\documentclass[useAMS,usenatbib]{mnras}
\usepackage{graphicx}
\usepackage{epstopdf}
\usepackage{amsmath,esint,bm}
\usepackage{amssymb}

\usepackage{etoolbox}
\makeatletter
\patchcmd\@combinedblfloats{\box\@outputbox}{\unvbox\@outputbox}{}{%
   \errmessage{\noexpand\@combinedblfloats could not be patched}%
}%
 \makeatother

\title[Potential and Lagrangian in Modified Gravity with Varying G]{Gravitational Potential and Nonrelativistic Lagrangian in Modified Gravity with Varying G}
\author[Christodoulou \& Kazanas]{Dimitris M. Christodoulou$^{1,2}$  and Demosthenes Kazanas$^{3}$
\\
$^{1}$Lowell Center for Space Science and Technology, University of Massachusetts Lowell, Lowell, MA, 01854, USA.\\
$^{2}$Dept. of Mathematical Sciences, Univ. of Massachusetts Lowell,
Lowell, MA, 01854, USA. E-mail: dimitris\_christodoulou@uml.edu\\
$^{3}$NASA/GSFC, Laboratory for High-Energy Astrophysics, Code 663, Greenbelt, MD 20771, USA. E-mail: demos.kazanas@nasa.gov\\
}

\begin{document}

\def\gsim{\mathrel{\raise.5ex\hbox{$>$}\mkern-14mu
                \lower0.6ex\hbox{$\sim$}}}

\def\lsim{\mathrel{\raise.3ex\hbox{$<$}\mkern-14mu
               \lower0.6ex\hbox{$\sim$}}}

%\date{Accepted ??? . Received 2017 March ??; in original form 2017 March ??}
\pagerange{\pageref{firstpage}--\pageref{lastpage}} \pubyear{2018}

\maketitle

\label{firstpage}

\begin{abstract}
We have recently shown that the baryonic Tully-Fisher (BTF) and Faber-Jackson (BFJ) relations imply that the gravitational ``constant'' $G$ in the force law vary with acceleration $a$ as $1/a$. Here we derive the converse from first principles. First we obtain the gravitational potential for all accelerations and we formulate the Lagrangian for the central-force problem. Then action minimization implies the BTF/BFJ relations in the deep MOND limit as well as weak-field Weyl gravity in the Newtonian limit. The results show how we can properly formulate a nonrelativistic conformal theory of modified dynamics that reduces to MOND in its low-acceleration limit and to Weyl gravity in the opposite limit. An unavoidable conclusion is that $a_0$, the transitional acceleration in modified dynamics, does not have a cosmological origin and it may not even be constant among galaxies and galaxy clusters.
\end{abstract}




\section{Introduction}\label{intro}

In previous work \citep{chr18}, we showed that in the regime in which the observed baryonic Tully-Fisher (BTF) \citep{tul77,mcg00,mcg12} and Faber-Jackson (BFJ)  \citep{fab76,san09,den15} relations are valid, the gravitational ``constant'' $G$ should vary with acceleration $a$ in the force law. Such a varying $G(a)$ function can naturally account for the non-Newtonian force postulated in Modified Newtonian Dynamics (MOND) \citep{mil83a,mil83b,mil83c,mil15a,mil15b,mil16}, as well as for additional terms that appear only in weak-field Weyl gravity \citep{man89,man94}.

In the deep MOND limit of $a\ll a_0$, where $a_0$ is a transitional acceleration and $G=G_0 a_0/a$, the variation of $G$ introduces only one universal constant, the product $G_0 a_0$ \citep[see also][]{mil15c}. Furthermore, the Weak Equivalence Principle remains valid since the inertial mass is not modified, whereas the Strong Equivalence Principle is invalid since $G$ varies at all scales. These findings suggest that $a_0$ may not have a cosmological origin despite the well-known numerical coincidence that $a_0\simeq cH_0\simeq 1.2\times 10^{-10}$~m~s$^{-2}$, where $c$ is the speed of light and $H_0$ is the Hubble constant. In fact, $G_0$ and $a_0$ may not even be constants among galaxies or clusters of galaxies; they may vary in space in a way that maintains their universal constant product.

This last statement may not be entirely clear: We measure $G_0$ in the laboratory at high accelerations and the measured value works well at solar-system scales. But our value of $G_0$ for $a\gg a_0$ is not independently constrained by a relation such as the BTF/BFJ relations for $a\ll a_0$. So we have no independent evidence that $G_0$ takes the same value at the center of our Galaxy or in other galaxies for that matter. This has become a major point of contention recently and we will return to it in \S~\ref{conc}.

Our previous work relied on important galaxy observations \citep{fab76,tul77} to establish a theoretical result, namely, that $G\propto 1/a$ at very low accelerations.
In this work, we demonstrate that the converse is also true and that it effectively ties up the existence of the BTF/BFJ relations to a single fundamental assumption, the variation of $G(a)$ in the force law. We formulate our derivations by obtaining the gravitational potential and the associated nonrelativistic Lagrangian of the central-force problem with varying $G(a)$ and then by considering the radial Euler-Lagrange equation in spherical symmetry. 

In \S~\ref{acc}, we derive the gravitational potential and the Lagrangian in the general case that includes the asymptotic cases as well as the intermediate accelerations regime. In \S~\ref{derivation}, we derive the BTF/BFJ relations and their first-order corrections as a special case in which $a\ll a_0$; as well as weak-field Weyl gravity as a correction term to Newtonian gravity in the Newtonian limit $a\gg a_0$. In \S~\ref{conc}, we discuss our results in light of the latest developments in the field.

\section{Gravitational Potential and Lagrangian in the Central-Force Problem with Varying G}\label{acc}

In the general case, applicable to all accelerations irrespective of magnitude $a$, the function $G(a)$ is given by the equation
\begin{equation}
G(a) = G_0 + \frac{G_0a_0}{a}\, ,
\label{xG1}
\end{equation}
where $G_0 = 6.674\times 10^{-11}$ m$^3$ kg$^{-1}$ s$^{-2}$ is the Newtonian value of the gravitational constant and $G_0a_0 = 8.0\times 10^{-21}$ m$^4$ kg$^{-1}$ s$^{-4}$ is a new characteristic constant that appears in the deep MOND limit \citep{mil15c}. This new constant has dimensions of $[v]^4/[M]$, a strong hint that it is related to the BTF/BFJ relations (see \S~\ref{derivation} below).

The two terms in eq.~(\ref{xG1}) are mandatory in order for $G(a)$ to describe correctly the gravitational force in the above two asymptotic cases. These terms combine to also describe the regime of intermediate accelerations. One might be tempted to use different functional forms for $G(a)$ with the appropriate limiting behaviors, as was also done for MOND with its arbitrary interpolating functions, but such different forms introduce additional spurious physics in the intermediate regime. For this reason, adoption of eq.~(\ref{xG1}) affords us less freedom in modifying the force law as compared to MOND whose dynamics depends only on the asymptotic form of the force and treats the intermediate regime as free of additional constraints.

\subsection{Gravitational Potential}\label{fb}

When the force law $a=G(a)M/r^2$ is modified by the varying $G(a)$, the gravitational potential $\Phi(r)$ of a central mass $M$ at distance $r$ is no longer equal to its Newtonian form $G(a)M/r$. Here we derive $\Phi(r)$ from the acceleration $a$ by integrating the equation 
\begin{equation}
a \equiv - \frac{d}{dr}\Phi(r)\, ,
\label{f1}
\end{equation}
where $a$ is derived from force balance \citep{chr18}, viz.\footnote{Eq.~(\ref{f2}) happens to be one of MOND's interpolating functions \citep[the ``simple'' function; e.g.,][]{fam12} and agrees very well with the empirical results of \cite{mcg16} and \cite{lel17} who measured the acceleration at $\sim$3000 distinct points in 153 and 240 galaxies, respectively.}
\begin{equation}
a = \frac{a_N}{2}\left(1 + \sqrt{1 + \frac{4a_0}{a_N}}\right) \, ,
\label{f2}
\end{equation}
where the Newtonian acceleration $a_N$ is defined by
\begin{equation}
a_N \equiv \frac{G_0M}{r^2}\, .
\label{f3}
\end{equation}
In eq.~(\ref{f1}), $\Phi(r)$ is defined implicitly without the customary negative sign so that the magnitudes of the accelerations will be strictly positive, viz. $a > 0$ and $a_N > 0$.

Substituting eqs.~(\ref{f2}) and~(\ref{f3}) into eq.~(\ref{f1}) and carrying out the integration over $r$, we find that
\begin{equation}
\frac{\Phi(x)}{\sqrt{G_0 M a_0}} = \frac{1+\sqrt{1+4x^2}}{2x} - \ln\left(2x + \sqrt{1 + 4x^2}\right) \, ,
\label{xf2}
\end{equation}
where the dimensionless radius $x$ is defined by
\begin{equation}
x \equiv r/r_M \, ,
\label{x}
\end{equation}
and the MOND characteristic radius $r_M$ is given by 
\begin{equation}
r_M = \sqrt{G_0 M/a_0} \, .
\label{rm}
\end{equation}

In the Newtonian limit $x\to 0$, eq.~(\ref{xf2}) reduces to $\Phi(r) \approx G_0M/r - a_0 r$ and the acceleration (eq.~(\ref{f1})) then is $a\approx a_N + a_0$. The Newtonian term $a_N$ was expected, whereas the non-Newtonian constant term has only been predicted in the weak-field limit of conformal Weyl gravity \citep{man89,man94}.

In the deep MOND limit $x\to\infty$, eq.~(\ref{xf2}) reduces to $\Phi(r) \approx -\sqrt{G_0 M a_0}\ln r + G_0M/(2r)$ and the acceleration (eq.~(\ref{f1})) then is $a\approx \sqrt{a_N a_0} + a_N/2$ \citep[see also][]{chr18}.

\subsection{Lagrangian Formulation}\label{lf}

The Lagrangian of a test particle orbiting around mass $M$ at distance $r$ is written in polar coordinates $(r, \theta)$ as
\begin{equation}
{\cal L}(r, v) = \frac{1}{2}v^2 - \Phi(r)\, ,
\label{L1}
\end{equation}
where the orbital speed $v=r(d\theta/dt)$, $t$ is the time, and the potential $\Phi$ (eq.~(\ref{xf2})) is written again without the negative sign to ensure that $a>0$.
An alternative form can be produced by using the constant specific angular momentum of the test particle $\ell = r\,v$ to eliminate $v$, but the calculations are actually easier when using eq.~(\ref{L1}).

The radial Euler-Lagrange equation for ${\cal L}(r, v)$ is
\begin{equation}
\frac{\partial{\cal L}}{\partial r}  - \frac{d}{dt}\left(\frac{\partial{\cal L}}{\partial v}\right) = 0\, .
\label{AL2}
\end{equation}
Substituting eqs.~(\ref{xf2})-(\ref{L1}) into eq.~(\ref{AL2}), we find again eq.~(\ref{f2}). This demonstrates that action minimization is consistent with the force-balance calculation.

\section{Asymptotic Forms}\label{derivation}

In \S~\ref{fb}, we found the following asymptotic expressions for the acceleration $a$:
\begin{itemize}
\item[(a)]{\it Newtonian limit} ($a\gg a_0$): $a\approx a_N + a_0$.
The constant $a_0$ amounts to a small correction to the Newtonian acceleration $a_N$ (eq.~(\ref{f3})). Such a constant deviation from Newtonian dynamics has not been tested yet in the Cassini mission data \citep[][fitted only a quadrupolar correction to the Cassini data]{hee14}.\\

\item[(b)]{\it Deep MOND limit} ($a\ll a_0$): $a\approx \sqrt{a_N a_0} + a_N/2$. Using force balance $a=v^2/r$ and eqs.~(\ref{f3}) and~(\ref{rm}), this approximation takes the form
\begin{equation}
\frac{v^4}{M} \approx G_0 a_0\left[1 + \frac{r_M}{r} + \frac{1}{2}\left(\frac{r_M}{r}\right)^2\right] \, .
\label{L4}
\end{equation}
This equation represents the observed BTF/BFJ relations in which $v^4= G_0 M a_0$ to zeroth order in $1/r$. It also shows that the quotient $v^4/M$ is indeed related to MOND's universal constant $G_0a_0$ which has precisely the same dimensions. The higher-order correction terms decrease with distance $r$, thus they become negligible at large scales.

\end{itemize}

\section{Discussion}\label{conc}

We have shown that a varying gravitational ``constant'' $G(a)\propto 1/a$ in the radial Euler-Lagrange equation of the central force problem of Newtonian mechanics implies the observed BTF/BFJ relations at very low accelerations $a$. We have also derived a solution for the gravitational potential of a point-mass (eq.~(\ref{xf2})) and the acceleration of an orbiting test particle valid for all acceleration regimes (eq.~(\ref{f2})) using the same Lagrangian formulation (\S~\ref{lf}).

The adopted function for $G(a)$ has a unique form (eq.~(\ref{xG1})) that describes correctly the behavior of galactic stellar kinematics in the two asymptotic regimes (Newtonian and MOND). The same form also describes the intermediate regime in which we do not introduce any additional physics by avoiding the use of more complicated $G(a)$ functions with the same asymptotic behaviors. In addition, each of the two asymptotic terms in eq.~(\ref{xG1}) generates a small contribution in the opposite limit: the Newtonian  term $a_N/2$ modifies MOND's acceleration, whereas the term $a_0$ modifies the Newtonian acceleration (\S~\ref{derivation}). Such a small non-Newtonian constant has been predicted in the weak-field limit of conformal Weyl gravity \citep{man89,man94}.  

It has been argued that a small quadrupolar correction to the Newtonian gravitational field of our solar system, obtained from Cassini monitoring radio data, is consistent with relativistic deviations at large solar-system scales and offers no support for MOND-type deviations \citep{hee14}. In that investigation, a quadrupolar term was actually fitted to the Cassini data and its magnitude was estimated. These results are not applicable to our case, where the correction to the Newtonian acceleration within the solar system is just a small constant term $a_0\sim 1\, {\rm \AA}\,{\rm s^{-2}}$ (\S~\ref{derivation}). This correction is produced by a linear potential of the form $\delta\Phi(r)=+a_0\,r$. The Cassini data will have to be fitted again for this potential, although it may be difficult for the analysis to detect a correction as small as $a_0$ ($a_0/a_N \simeq 2\times 10^{-6}$ at the distance of Saturn).

The above formulation of modified dynamics with $G(a)$ given by eq.~(\ref{xG1}) shows that the only constant introduced in the deep MOND limit is $G_0a_0$. This unusual constant was already known to \cite{mil15c} who argued that the product maintains scale invariance in MOND. But this is a mathematical argument and it implies that $a_0$ is not necessarily a constant of MOND in its deep limit. On the other hand, $a_0$ appears alone as a constant only in the Newtonian regime of accelerations, where $a \approx a_N + a_0$ (\S~\ref{derivation}). In our modified dynamics, the new constant is introduced by the varying $G(a)$ itself. As such, the term $G_0a_0/a$ does not have an obvious cosmological underpinning, it is rather localized to large scales in individual galaxies and it is in fact responsible for the appearance of the small Weyl-like correction $a_0$ to the acceleration in the Newtonian regime. Furthermore, it remains an open question whether $G_0$ and $a_0$ are separate constants among individual galaxies and clusters of galaxies. At such large scales, the individual values could possibly vary in a way that maintains a constant universal product $G_0a_0$.

Recently, \cite{rod18} argued that $a_0$ cannot be constant in individual galaxies whose rotation curves were used to obtain its best-fit value. On a statistical basis, a constant $a_0$ was rejected at more than the 10$\sigma$ level of significance. This appears to be a much stronger result than from previous studies \citep{ran14,ioc15,hee16} which also indicated that $a_0$ may not be constant between different galaxies.
\cite{rod18} concluded that MOND is not a viable theory on galactic scales. This conclusion is premature and it has already been disputed forcefully in the published literature \citep{li18}. If the above studies are confirmed by future independent investigations, the results may constitute evidence that $G_0$ and $a_0$ vary from galaxy to galaxy in a way that their product remains a universal constant. If true, such behavior would make the study of gravitation in galaxies and galaxy clusters a lot more complicated.

%\cite{dip18} have just announced new empirical evidence that the radial acceleration relation (RAR) found by \cite{mcg16} and \cite{lel17} is only an upper limit of the deviations of the observed gravity from its Newtonian value. In particular, they obtained 1601 measurements of the radial acceleration in dwarf and low surface brightness (LSB) galaxies and they found that the deviations depend strongly on the radii where the measurements are made. The measurements across the LSB galaxies at relatively small radii stand out in that they deviate litlle from the Newtonian values or not at all. \cite{dip18} concluded that the RAR depends on an additional independent parameter, the galactic radius where the measurement is performed. However, there is another explanation. These results can make sense if $G_0$ and $a_0$ vary individually between galaxies. Then in small galaxies, there is not enough radial extent for $G(a)$ to increase substantially with radius and the accelerations remain close to their Newtonian values.

\section*{Acknowledgments}
We thank an anonymous referee for a thoughtful review of this paper.
NASA support over the years is gratefully acknowledged.

\label{lastpage}

\end{document}